# Laser spectroscopy of individual quantum dots charged with a single hole

B. D. Gerardot, R. J. Barbour, D. Brunner, and P. A. Dalgarno
*School of Engineering and Physical Sciences, Heriot-Watt University, Edinburgh EH14 4AS, UK*

A. Badolato
*Department of Physics and Astronomy, University of Rochester, Rochester, New York 14627, USA*

N. Stoltz and P. M. Petroff
*Materials Department, University of California, Santa Barbara, California 93106, USA*

J. Houel and R. J. Warburton
*Department of Physics, University of Basel, Klingelbergstrasse 82, CH-4056 Basel, Switzerland*

We characterize the positively charged exciton ($X^{1+}$) in single InGaAs quantum dots using resonant laser spectroscopy. Three samples with different dopant species (Be or C as acceptors, Si as a donor) are compared. The p-doped samples exhibit larger inhomogeneous broadening (x3) and smaller absorption contrast (x10) than the n-doped sample. For $X^{1+}$ in the Be-doped sample, a dot dependent non-linear Fano effect is observed, demonstrating coupling to degenerate continuum states. However, for the C-doped sample the $X^{1+}$ lineshape and saturation broadening follows isolated atomic transition behaviour. This C-doped device structure is useful for single hole spin initialization, manipulation, and measurement.

Single spins in semiconductor quantum dots (QDs) are promising for applications which exploit quantum superposition[1]. Recent advances demonstrate the potential of a single electron trapped in QDs for such purposes[2]. However, the hyperfine interaction of the electron spin with the fluctuating nuclear spin bath leads to dephasing and presents considerable challenges. One approach to reduce or even avoid these challenges is to use valence-band holes instead of electrons as quantum bits[3-12]. Crucially, although a QD hole still experiences a hyperfine interaction with a coupling coefficient ~ 10% that of an electron spin[9-11], the hyperfine interaction for an ideal heavy hole takes on an Ising-like form such that heavy hole spin dephasing is greatly suppressed by an in-plane magnetic field[12].

Recent experiments report a range of hole spin ensemble coherence times ($T_2^*$), from a few ns up to hundreds of ns[5-7]. For the experiments yielding smaller dephasing times, the hyperfine interaction is not the limiting dephasing mechanism[6,7]. An observation that highlights the importance of the sample, device and experimental design is that charge noise may also translate into spin dephasing via the dependence of the hole g-factor on electric field. Motivated by this, we probe here positively charged excitons in field effect devices. We exploit laser spectroscopy which gives information on exciton dephasing via the linewidth. Furthermore, the lineshape is a powerful diagnostic of the nature, closed or open, of the quantum system as even weak couplings to a degenerate continuum lead to departures from a simple Lorentzian. We uncover a clear advantage of C-doping over Be-doping.

In general, there are three approaches to load holes into a quantum dot: modulation doping near the QD layer[11], optical generation[3,8,13], or controlled tunnelling from a nearby Fermi sea[4-7,14]. We employ the last of these methods and study three samples all similar in structure except for the dopant species in the back gate. Two samples contain different acceptor dopants: Be or C. These are compared to a well characterized sample containing Si as a donor dopant [same sample as Refs. [15-17]]. The samples, grown by molecular beam epitaxy, consist of dots tunnel coupled to a grounded back contact (doping ~ 4 x $10^{18}$ $cm^{-3}$) through a 25 nm barrier. The dots are capped with a GaAs layer of thickness *d*, an AlAs / GaAs superlattice with thickness (130 – *d)* nm, a 6 nm GaAs layer, and finally a 4 nm semi-transparent NiCr Schottky gate to which a bias is applied. For the n (p)-type sample(s), *d* = 10 (30) nm.

To investigate the charging behavior we perform initially non-resonant photoluminescence (PL) spectroscopy on single dots for each sample at 4.2 K using a confocal microscope. The PL is excited with an 830 nm laser diode, dispersed with a grating spectrometer, and detected with a Si charge coupled device. Figs. 1(a) and (b) show contour plots of the luminescence as a function of applied bias for the Si-doped and C-doped samples, respectively. The changes in emission energy as a carrier is added arises from the Coulomb interaction, which is well characterized for each charge configuration allowing straightforward assignment of each spectral line[17]. For the Si-doped sample, a single charge configuration is dominant at any one gate bias, Fig. 1(c). For the C-doped sample, multiple charge states with similar intensities appear in the spectra at a single bias, Fig. 1(d). In general, the appearance of multiple emission lines can arise with non-resonant excitation in time-integrated spectra[18]. Here, the different charging behaviour observed in the PL spectra is a consequence of the different carrier tunnelling times ($\tau_{tun}$) in the two devices, which we have previously determined to be ~ 10 ps and 10 ns for the Si- and C- doped samples with 25 nm tunnel barriers, respectively[4,19]. For the n-type sample, $\tau_{tun} << \tau_{rad}$ (radiative lifetime) for the $X^{1+}$ (1.2 ns), $X^0$ (0.8 ns), and $X^{1-}$ (0.9 ns)[17]. This ensures that the exciton relaxes to its ground-state via an interaction with the Fermi sea within its lifetime. However, for the p-type sample $\tau_{tun} > \tau_{rad}$ and the exciton may or may not relax within its lifetime, creating a large probability of overlapping lines in the time-averaged spectra.

Once the QD states are characterized in PL, we use a much more refined technique, resonant laser spectroscopy with differential transmission detection[20], to characterize the true lineshape and linewidth of the exciton transitions. A glass ($n$ = 2.0 for $\lambda$ = 950 nm) hemispherical solid immersion lens on top of each sample is used to reduce the spatial resolution and increase the signal:noise[15]. Unlike non-resonant excitation, the resonant laser is spectrally specific, interacting at any one time only with one charge configuration. Fig. 1(e) demonstrates well defined Coulomb blockade for charging of a single hole in a QD. Compared to PL, with resonant excitation the $X^0$ transition persists to much more positive voltages. However, for increasingly large positive voltages the hole tunnelling rates increase, increasing the $X^0$ linewidth and decreasing the transmission contrast[21].

Figs. 2(a) and (b) show typical differential transmission results for the $X^0$ transitions (at voltages where carrier tunnelling rates are negligible) with circularly polarized light from the Si- and C-doped samples, respectively. Due to the electron-hole exchange interaction, the $X^0$ transition splits into two linearly polarized transitions[20]. There are two striking differences between the n- and p-type devices: the signal strength at resonance (the contrast) is ~ 10x smaller and the transition linewidths are ~ 3x larger for the C-doped compared to the Si-doped sample. We find similar $X^0$ linewidths for the C- and Be-doped structures. The typical linewidth for $X^0$ if it were determined only by spontaneous emission corresponds to ~ 0.8 μeV (dashed lines in Fig. 2(b) and (d)), but the linewidths observed in these particular devices are all significantly larger. The inhomogeneous broadening is likely caused by spectral fluctuations, slow on the timescale determined by radiative recombination, but fast compared to the measurement integration rate[4,20]. No correlation between the measured linewidth and Stark shift is observed and the experimental lineshapes are best fit by a pure Lorentzian function (instead of a composite Lorenztian-Gaussian function). While the different capping layer thicknesses for the n- and p-type samples may contribute to the disparity in linewidths, the difference in contrast is presently not understood.

Fig. 3 shows typical transmission spectra for singly charged QDs from the Si-, Be-, and C-doped samples at zero applied magnetic field. In each case linear polarization was used to prevent optical pumping[4]. Similar to the transmission spectra for the neutral excitons in Fig. 2, we find that the $X^{1-}$ transitions in the Si-doped sample and the $X^{1+}$ transitions in the C-doped sample display Lorentzian lineshapes. Further evidence of a closed atomic transition is the manifestation of near ideal saturation broadening (see ref. [15] for the $X^0$ and $X^{1-}$ saturation curves from the Si-doped sample). However, the differential transmission spectra for the $X^{1+}$ transitions in the Be-doped sample (Fig. 3(b) and Fig. 5(a-e)) display dispersive-like lineshapes with positive overshoots and crossings through the 100% line. Additionally, the $X^{1+}$ transitions saturate at much lower power than the $X^0$ on the same dot and therefore do not follow the two-level model (see Fig. 4). Furthermore, for some dots, the lineshape changes as a function of power (see Fig. 5). These results point to the observation of a non-linear Fano effect[22] on the $X^{1+}$ in the Be-doped sample. An explanation of the dispersive lineshapes in terms of an optical interference effect[23,24] can be ruled out from the sample-independent dot-to-surface distance, from the close-to-Lorentzian lineshapes expected in transmission geometry[24] and in particular from the power dependence of the lineshape.

The Fano effect, a detuning-dependent quantum interference between an atom-like and continuum transition[25], is parameterized by the Fano factor $q$. $q \rightarrow \infty$ corresponds to negligible interference in the atom-like transition with the continuum, resulting in a Lorenztian lineshape; conversely, $q = 1$ signifies a strong quantum interference with the nature, destructive or constructive, changing rapidly with detuning, leading to a dispersive lineshape. In Fig. 3(b) [5(a)], we find an excellent fit to the Fano lineshape with $q = 1.2$ [$q = 3.9$]. The recent insight into Fano physics using quantum dots is that for modest values of $q$, the visibility of the Fano interference increases with laser power once the two-level transition saturates[22], an apparent decrease in $q$ with laser power. This is exactly the behaviour we observe here, Fig. 5(a)-(e), adding considerably weight to our assertion that the non-Lorenztian lineshapes for $X^{1+}$ in the Be-doped sample arises from a Fano effect.

The key difference between the Be-doped sample and the other samples is the dopant, suggesting that impurity states from Be are the source of continuum states. For MBE growth of Be doped [001] GaAs, it is difficult to grow sharp interfaces due to significant Be diffusion at typical growth temperatures[26]. Alternatively, C doping with a $CBr_4$ source offers the ability to obtain high doping concentrations and much sharper interfaces in high quality GaAs[27]. The $X^{1+}$ transmission spectra therefore suggest that Be dopant atoms form an electronic continuum at the $X^{1+}$ energy, within the GaAs gap. The microscopic mechanism is unclear and, to the best of our knowledge, such hybridized, Be-related deep levels have not been previously observed. We stress the power of the technique: the weak, broadband continuum transition is rendered visible by interfering it with the strong, narrow quantum dot transition. The fluctuating dot environment is demonstrated by the varying Fano visibility. For instance, QD1 shows a smaller $q$ than QD3: $q$=1.0 regardless of excitation power, Fig. 5(f).

The results show that, in general, the inhomogeneous linewidth is larger, the absorption contrast smaller, and the carrier tunnelling times larger for samples with *p*-doped back gates compared to *n*-doped. Nevertheless, resonant laser spectroscopy reveals ideal Coulomb blockade for *p*-doped structures. A dot dependent non-linear Fano effect is observed for $X^{1+}$ transitions from the Be-doped sample. This points to a coupling to continuum states likely caused by defect states created by Be dopant atoms which have diffused towards the QD layer. This continuum coupling is undesirable for an application as a spin qubit. However, laser spectroscopy on the $X^{1+}$ transition in the sample with C-doping reveals near ideal two-level behaviour. This sample structure is promising for probing and exploiting hole spin coherence.

The authors acknowledge the financial support for this work from EPSRC, the Royal Society, the Royal Society of Edinburgh and NCCR QSIT.

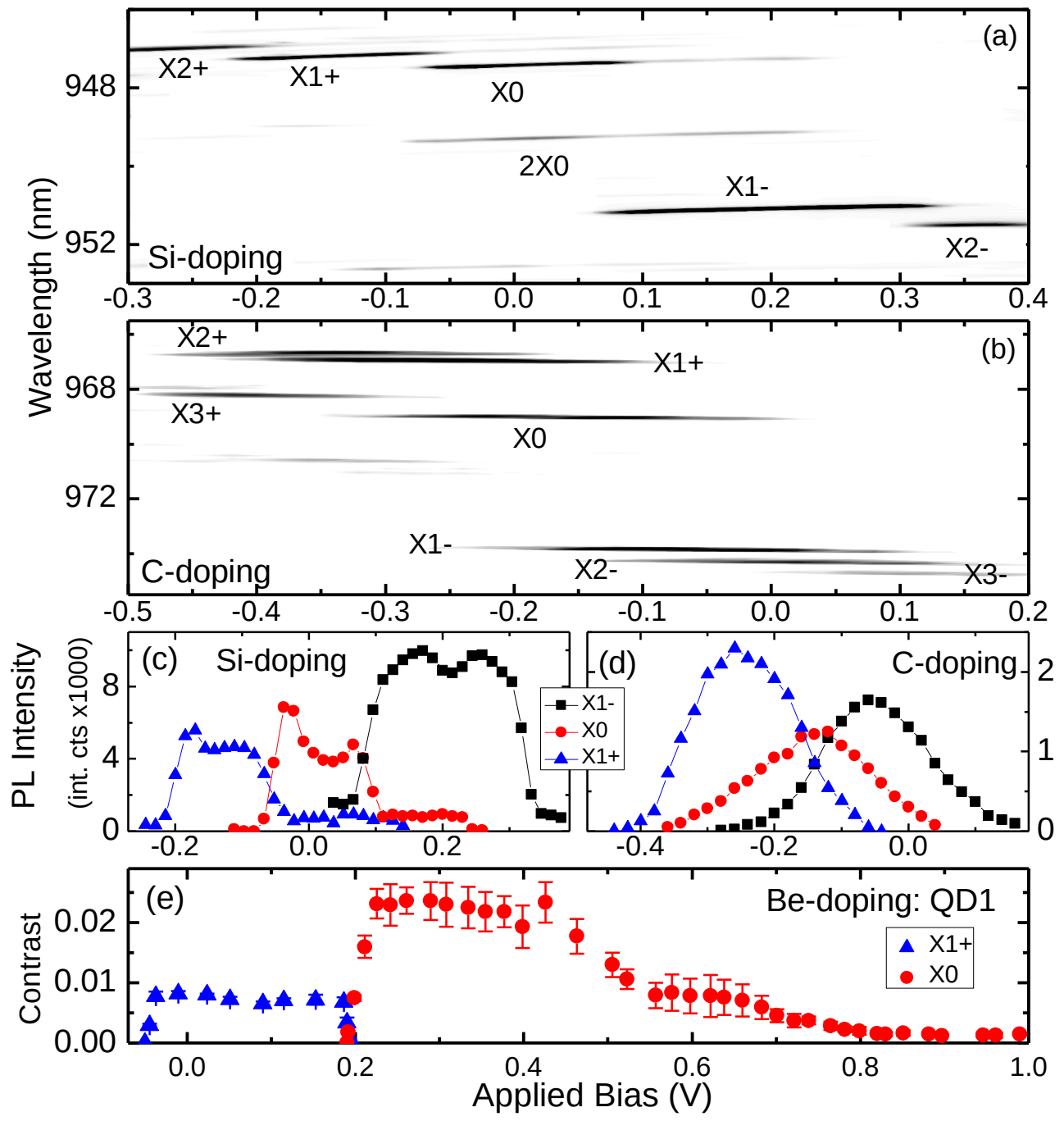

**Figure 1**: (color online) Grey-scale intensity contour plots of the collected photon wavelength as a function of applied bias for single dots in the (a) Si- and (b) C-doped sample using 10 and 20 s integration times, respectively. The integrated intensity for $X^{1-}$, $X^{0}$ and $X^{1+}$ at each applied bias for the Si- and C-doped samples are shown in (c) and (d), respectively. (e) The differential transmission contrast for $X^{1+}$ and $X^{0}$ for QD1 in the Be-doped sample. A resonant laser power of 0.4 nW was used with linear polarization so that only one state of the $X^{0}$ fine-structure is excited. As shown in Fig. 4, this driving power is still greater than the saturation power for this $X^{1+}$ transition.

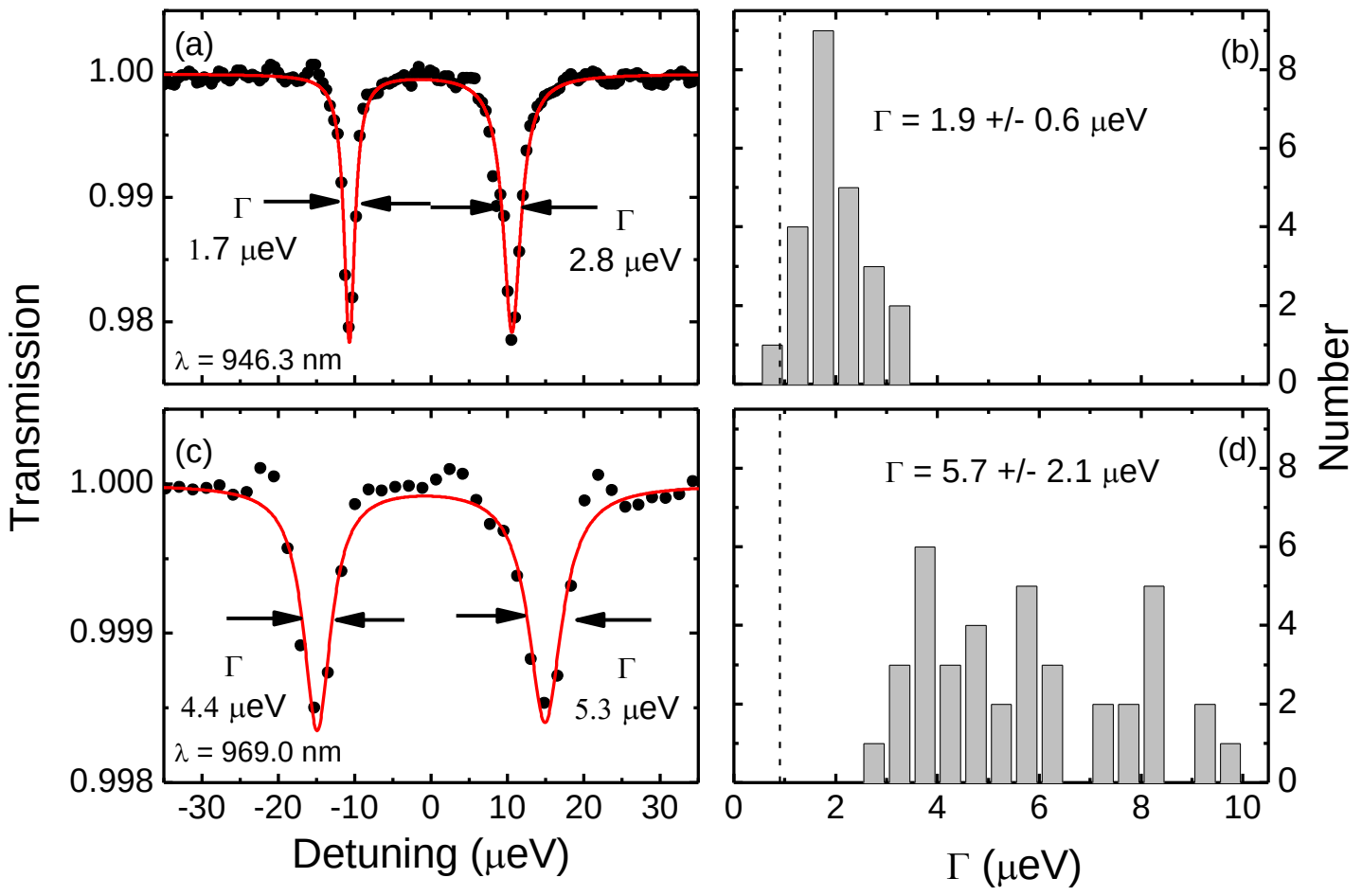

**Figure 2**: (color online) Typical transmission spectra (T = 4 K) using a resonant laser power of 0.08 nW and circular polarization for the $X^0$ transitions from the Si-doped (a) and C-doped (c) samples. Linewidth (full-width-at-half-maxium) statistics for each sample are shown in the histograms (b) and (d).

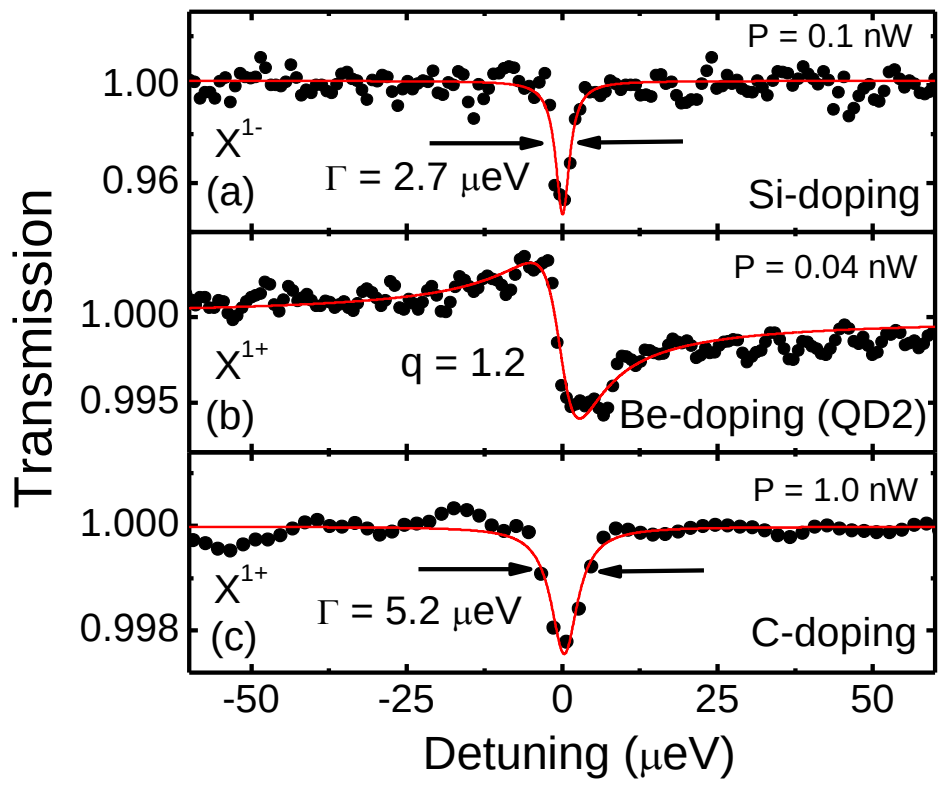

**Figure 3**: (color online) Typical spectra using linearly polarized resonant excitation for charged excitons from the 3 samples.

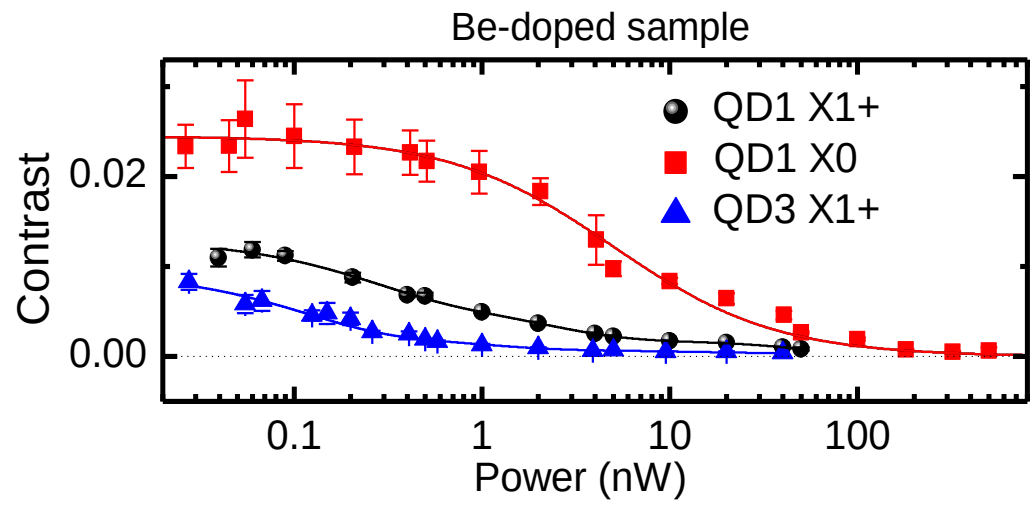

**Figure 4**: (color online) (a) Saturation curves for the $X^0$ and $X^{1+}$ transitions from QD1 and the $X^{1+}$ transition from QD3. The solid line for QD1 $X^0$ is a fit (with $\alpha_0$ = 0.0245, linewidth = 3.2 μeV, and E = 1.3266 eV) based on saturation of a two-level transition[15,24]. The solid lines for the $X^{1+}$ data are guides to the eye.

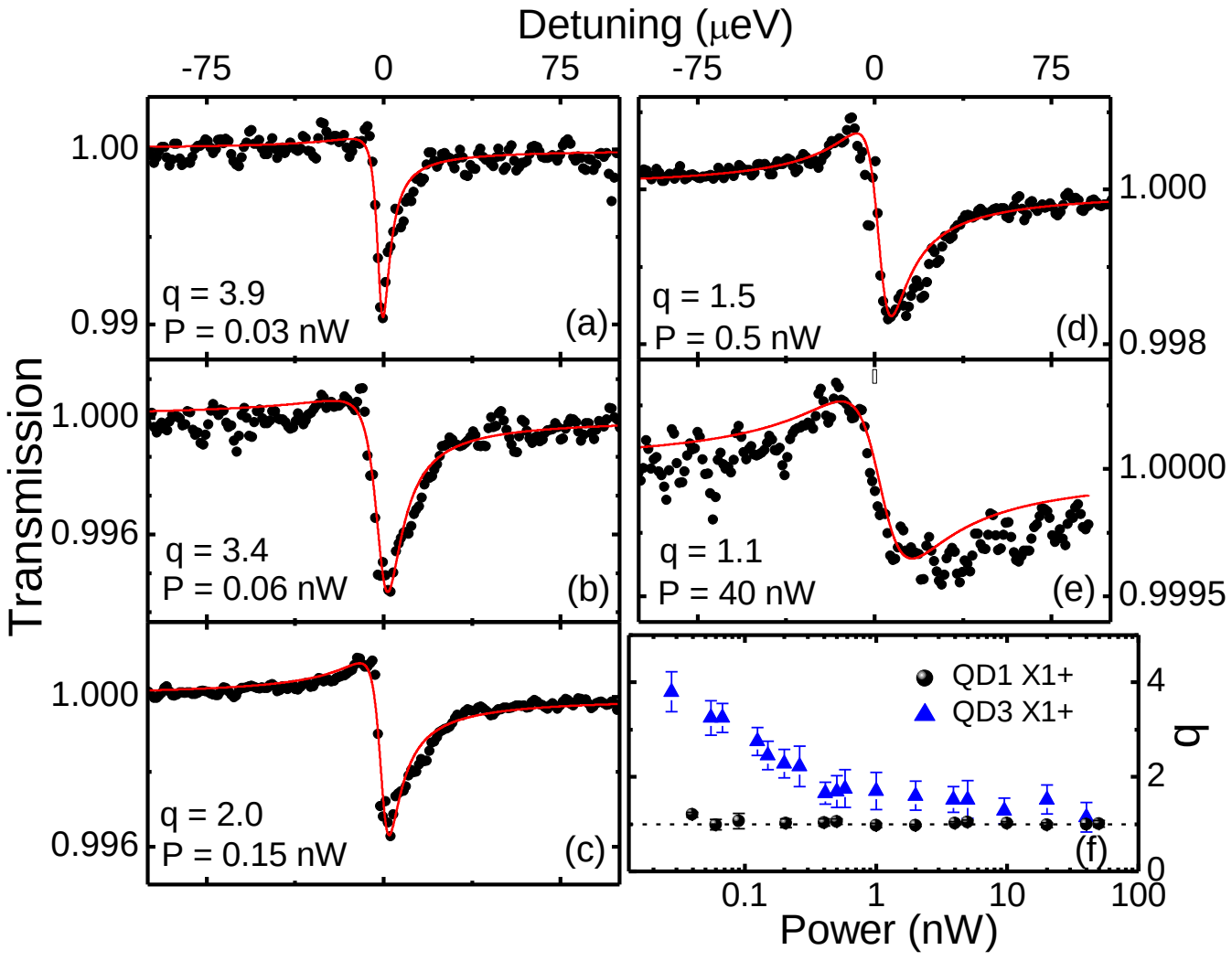

**Figure 5**: (color online) The evolution of the lineshape as a function of excitation power (a-e) for QD3 $X^{1+}$ from the Be-doped sample. The solid lines are fits using *q* as a free parameter. The Fano visibility as a function of power for this dot and also QD1 $X^{1+}$ is shown in (f).